\RequirePackage{lineno}
\documentclass[conference]{IEEEtran}

\usepackage{threeparttable}
\usepackage{lipsum}
\usepackage[dvips]{color}
\usepackage{lipsum}
\usepackage{algorithm,algorithmic}
\usepackage{epsfig,amsmath,amsfonts,cite}
\usepackage{booktabs}
\newcommand{\thickhline}{\noalign{\hrule height 1.0pt}}
\usepackage{multirow}

\newcommand{\ten}[1]{\mathbfcal{#1}}   

\DeclareMathAlphabet\mathbfcal{OMS}{cmsy}{b}{n}

\newcommand{\parm}{{\xi}}
\newcommand{\vecpar}{\boldsymbol{\parm}}

\newcommand{\parNum}{d}

\newcommand{\out}{y}

\newcommand{\multiGPC}{\Psi }

\newcommand{\polyInd}{\alpha}
\newcommand{\basisInd}{\boldsymbol{\polyInd}}

\newcommand{\pcOrder}{p}

\newcommand{\yPC}{\sum\limits_{|\basisInd|=0}^{\pcOrder} {c_{\basisInd}  \multiGPC_{\basisInd}  (\vecpar)} }

\newcommand{\uvec}{\mathbf{u} }

\begin{document}
%
\title{A Big-Data Approach to Handle Process Variations: Uncertainty Quantification by Tensor Recovery}


\author{\IEEEauthorblockN{Zheng Zhang, Tsui-Wei Weng and Luca Daniel}
\IEEEauthorblockA{Research Laboratory of Electronics, Massachusetts Institute of Technology, Cambridge, MA 02139, USA\\
E-mail: z\_zhang@mit.edu, twweng@mit.edu, luca@mit.edu}}


%


\maketitle

\begin{abstract}
Stochastic spectral methods have become a popular technique to quantify the uncertainties of nano-scale devices and circuits. They are much more efficient than Monte Carlo for certain design cases with a small number of random parameters. However, their computational cost significantly increases as the number of random parameters increases. This paper presents a big-data approach to solve high-dimensional uncertainty quantification problems. Specifically, we simulate integrated circuits and MEMS at only a small number of quadrature samples; then, a huge number of (e.g., $1.5 \times 10^{27}$) solution samples are estimated from the available small-size (e.g., $500$) solution samples via a low-rank and tensor-recovery method. Numerical results show that our algorithm can easily extend the applicability of tensor-product stochastic collocation to IC and MEMS problems with over 50 random parameters, whereas the traditional algorithm can only handle several random parameters.
\end{abstract}


%
\IEEEpeerreviewmaketitle

\section{Introduction}
Fabrication process variations can significantly decrease the yield of nano-scale chip design~\cite{variation2008}. In order to estimate the uncertainties of chip performance, Monte Carlo~\cite{MCintro, SingheeR09} has been used in commercial electronic design automation software for decades. Monte Carlo is easy to implement but generally requires a large number of repeated simulations due to its slow convergence rate. In recent years, stochastic spectral methods~\cite{sfem, col:2005} have emerged as a promising alternative due to their higher efficiency for certain design cases.

Stochastic spectral methods approximate a stochastic solution as a linear combination of some generalized polynomial chaos basis functions~\cite{gPC2002}, and they may get highly accurate solutions without (or with only a small number of) repeated simulations.  Based on intrusive (i.e., non-sampling) formulations such as stochastic Galerkin~\cite{sfem} and stochastic testing~\cite{zzhang:tcad2013} as well as sampling-based formulations such as stochastic collocation~\cite{col:2005},  extensive results have been reported for IC~\cite{manfredi:tcas2014, Stievano:2011_1, zzhang:tcad2013, Strunz:2008, zzhang:tcas2_2013,Pulch:2011_1, Rufuie2014}, MEMS~\cite{zzhang_cicc2014, zzhang:huq_tcad} and photonic~\cite{twweng:optsEx} applications. 

Unfortunately, the computational cost of stochastic spectral methods increases very fast as the number of random parameters increases. In order to solve high-dimensional problems, several advanced algorithms have been developed. For instance, analysis of variance (ANOVA)~\cite{zzhang_cicc2014} and compressed sensing~\cite{xli2010} can exploit the sparsity of high-dimensional generalized polynomial-chaos expansion. The dominant singular vector method~\cite{Tarek_DAC:10} can exploit the low-rank property of the matrix formed by all coefficient vectors. Stochastic model-order reduction~\cite{MoselhyD10} can remarkably reduce the number of simulations for sampling-based solvers. With devices or subsystems described by high-dimensional generalized polynomial-chaos expansions, hierarchical uncertainty quantification based on tensor-train decomposition~\cite{zzhang:huq_tcad} can handle more complex systems by changing basis functions.

This paper provides a big-data approach for solving the challenging high-dimensional uncertainty quantification problem. Specifically, we investigate tensor-product stochastic collocation that was only applicable to problems with a few random parameters. We represent the huge number of required solution samples as a tensor~\cite{tensor:suvey} (i.e., a high-dimensional generalization of matrix) and develop a low-rank and sparse recovery method to estimate the whole tensor from only a small number of solution samples. This technique can easily extend the applicability of tensor-product stochastic collocation to problems with over $50$ random parameters, making it even much more efficient than sparse-grid stochastic collocation. This paper aims at briefly presenting the key idea and showing its application in IC and MEMS. Theoretical and implementation details can be found in our preprint manuscript that is focused on power system applications~\cite{Zhang:tensor_ls15}.

\section{Preliminaries}
\subsection{Uncertainty Quantification using Stochastic Collocation}
Let $\vecpar=[\parm_1, \cdots, \parm_{\parNum}] \in \mathbb{R}^{\parNum}$ denotes a set of mutually independent random parameters that describe process variations. We aim to estimate the uncertainty of $\out (\vecpar)$, which is a parameter-dependent output of interest (e.g., the power consumption or frequency of a chip design). When $\out$ smoothly depends on $\vecpar$ and when it has a bounded variance, a truncated generalized polynomial-chaos expansion can be applied
\begin{equation}
\label{eq:ygpc}
\out (\vecpar) \approx \yPC, \; {\rm with}\; \mathbb{E}\left[{\multiGPC}_{\basisInd} \multiGPC_{\boldsymbol{\beta }}\left( \vecpar \right)\right ]=\sigma_{\basisInd, \boldsymbol{\beta }}.
\end{equation}
Here $\mathbb{E}$ denotes expectation, $\sigma$ denotes a Delta function, the basis functions $\{{\multiGPC}_{\basisInd} \left(\vecpar\right)\}$ are some orthonormal polynomials, $\basisInd=[\alpha_1,\cdots, \alpha_{\parNum}] \in \mathbb{N}^{\parNum}$ is a vector indicating the highest polynomial order of each parameter in the corresponding basis. The total polynomial order $|\basisInd|$ is bounded by $p$, and thus the total number of basis functions is $(p+d)!/(p!d!)$.

The coefficient $c_{\basisInd}$ can be obtained by a projection:
\begin{equation}
\label{eq:yproject}
c_{\basisInd}= 
\int\limits_{\mathbb{R}^d} {\out (\vecpar)  {\multiGPC}_{\basisInd} (\vecpar)\rho({\vecpar}) d\vecpar}, \;{\rm with}\; \rho(\vecpar)=\Pi_{k=1}^d {\rho _k(\xi _k)}
\end{equation}
where $\rho(\vecpar)$ is the joint probability density function and $\rho_k(\xi_k)$ the marginal density of $\xi_k$. The above integral needs to be evaluated with some numerical techniques. This paper considers the tensor-product implementation. Specifically, let $\{(\xi_k^{i_k}, w_k^{i_k})\}_{i_k=1}^q$ be $q$ pairs of quadrature points and weights for parameter $\xi_k$, the integral in (\ref{eq:yproject}) is evaluated as
\begin{equation}
\label{eq:yTP}
c_{\basisInd}=\sum\limits_{1\leq i_1,\cdots, i_d\leq q} {\out (\vecpar_{i_1\cdots i_d})  {\multiGPC}_{\basisInd}  (\vecpar_{i_1\cdots i_d})  w_{i_1\cdots i_d}  }
\end{equation}
where $\vecpar_{i_1\cdots i_d}=[\xi_1^{i_1}, \cdots, \xi_d^{i_d}]$ and $w_{i_1\cdots i_d}=w_1^{i_1}\cdots w_d^{i_d}$. Very often, evaluating each solution sample $\out (\vecpar_{i_1\cdots i_d})$ requires a time-consuming numerical simulation (e.g., a RF circuit simulation that involves periodic steady-state computation or a detailed device simulation by solving a large-scale partial differential equation or integral equation formulation).


\subsection{Tensor and Tensor Decomposition}
\begin{figure}[t]
	\centering
		\includegraphics[width=3.3in]{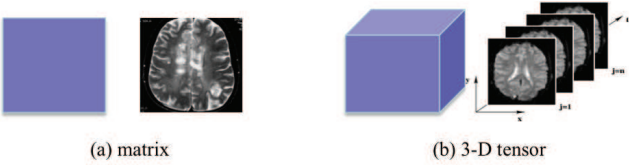} 
\caption{(a) a 2-D data array (e.g., a medical image) is a matrix, (b) a 3-D data array (e.g., multiple slices of images) is a tensor.}
	\label{fig:tensor}
\end{figure} 

\subsubsection{Tensor} Tensor is a high-dimensional generalization of matrix. A matrix $\mathbf{X} \in \mathbb{R}^{n_1\times n_2}$ is a $2$nd-order tensor, and its element indexed by $(i_1, i_2)$ can be denoted as $x_{i_1i_2}$. For a general $d$th-order tensor $\ten{X} \in \mathbb{R}^{n_1\times \cdots n_d}$, its element indexed by $(i_1, \cdots, i_d)$ can be denoted as $a_{i_1\cdots i_d}$. Fig.~\ref{fig:tensor} shows a matrix and  a $3$rd-order tensor. Given any two tensors $\ten{X}$ and $\ten{Y}$ of the same size, their inner product is defined as
\begin{equation}
\langle \ten{X}, \ten{Y} \rangle :=\sum\limits_{i_1\cdots i_d} {x_{i_1\cdots i_d} y_{i_1\cdots i_d}}.
\end{equation}
The Frobenius norm of tensor $\ten{X}$ is further defined as $|| \ten{X}||_F :=\sqrt{\langle \ten{X}, \ten{X} \rangle}$.

\subsubsection{Tensor Decomposition} 
\begin{figure}[t]
	\centering
		\includegraphics[width=3.0in]{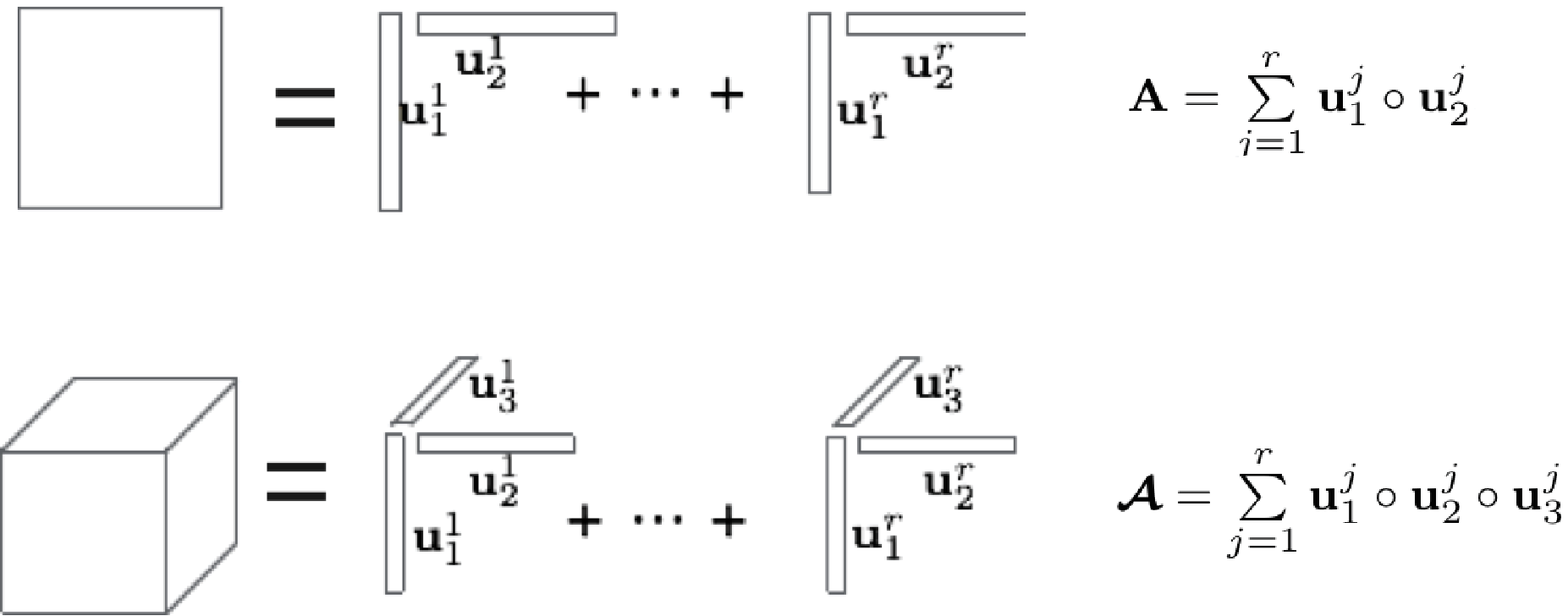} 
\caption{Low-rank factorization of a matrix (top) and of a $3$rd-order tensor (bottom).}
	\label{fig:tensor_fac}
\end{figure}

A tensor $\ten{X} $ is rank-1 if it can be written as the outer product of some vectors:
\begin{equation}
\label{eq:tensor_rank1}
\ten{X}=\uvec_1 \circ \cdots \circ \uvec_{\parNum}\; \Leftrightarrow\; x_{i_1\cdots i_{\parNum}}=\uvec_1(i_1) \cdots  \uvec_{\parNum}(i_{\parNum})
\end{equation}
where $\mathbf{u}_k(i_k)$ denotes the $i_k$-th element of vector $\mathbf{u}_k \in  \mathbb{R}^{n_k}$. Similar to matrices, a low-rank tensor can be written as the sum of some rank-1 tensors:
\begin{equation}
\label{eq:tensor_rank_r}
\ten{X}=\sum\limits_{j=1}^r {\uvec_1^j \circ \cdots \circ \uvec_{\parNum}^j}.
\end{equation}
As a demonstration, Fig.~\ref{fig:tensor_fac} shows the low-rank factorizations of a matrix and $3$rd-order tensor, respectively.

\section{Tensor Recovery Approach}
Formulation (\ref{eq:yTP}) is only applicable to problems with $5$ or $6$ random parameters due to the $q^d$ simulation samples required. This section describes our tensor-recovery approach that can significantly reduce the computational cost and extend (\ref{eq:yTP}) to design cases with many random parameters.

\subsection{Stochastic Collocation Using Tensor Recovery}

\subsubsection{Reformulation with Tensors} We define two tensors $\ten{Y}$ and $ \ten{W}_{\basisInd} $, such that their elements indexed by $(i_1, \cdots i_d)$ are $y(\vecpar_{i_1\cdots i_d })$ and ${\multiGPC}_{\basisInd}  (\vecpar_{i_1\cdots i_d})  w_{i_1\cdots i_d}$ respectively. Then, the operation in (\ref{eq:yTP}) can be written with tensors in a compact way:
\begin{equation}
c_{\basisInd} =\langle \ten{Y}, \ten{W}_{\basisInd} \rangle.
\end{equation}
It is straightforward to show that $\ten{W}_{\basisInd}$ is a rank-1 tensor. Our focus is to compute $ \ten{Y}$. Once $ \ten{Y}$ is computed, stochastic collocation can be done easily. Unfortunately, directly computing $\ten{Y}$ is impossible for high-dimensional cases, since it requires simulating a design problem $q^d$ times. 

\subsubsection{Tensor Recovery Approach} In order to reduce the computational cost, we estimate  $ \ten{Y}$ using an extremely small number of its elements. Let ${\cal I}$ include all indices for the elements of $\ten{Y}$, and its subset $\Omega$ includes the indices of a few available tensor elements obtained by circuit or MEMS simulation. With the sampling set  $\Omega$, a projection operator $\mathbb{P}$ is defined for $\ten{Y}$:
\begin{equation}
\label{tensor_project}
\ten{B}=\mathbb{P}_{\Omega}\left({\ten{Y}}\right) \; \Leftrightarrow\;  b_{i_1\cdots i_d} = \left\{ \begin{array}{l}
 y_{i_1\cdots i_d} ,\;{\rm{if}}\; {i_1\cdots i_d} \in  {\Omega} \\
 0 ,\;{\rm{otherwise}}.
 \end{array} \right.
 \end{equation}
We want to find a tensor $\ten{X}$ such that it matches $\ten{Y}$ for the elements specified by $\Omega$:
 \begin{equation}
\label{tensor_point_match}
\| \mathbb{P}_{\Omega}\left(\ten{X} -{\ten{Y}}\right)\|_F^2 =0.
\end{equation}
However, this problem is \textbf{ill-posed}, because any value can be assigned to $x_{i_1\cdots i_d}$ if $i_1\cdots i_d \notin \Omega$.

\subsubsection{Regularize Problem (\ref{tensor_point_match})} In order to make the tensor recovery problem well-posed, we add the following constraints based on practical observations.
\begin{itemize}
\item \textbf{Low-Rank Constraint:} we observe that very often the high-dimensional simulation data array $\ten{Y}$ has a low tensor rank. Therefore, we expect that its approximation $\ten{X}$ also has a low-rank property.  Thus, we assume that $\ten{X}$  has a rank-$r$ decomposition described in  (\ref{eq:tensor_rank_r}).

\item \textbf{Sparse Constraint:} as shown in practical cases~\cite{zzhang_cicc2014,xli2010}, most of the coefficients in a high-dimensional generalized polynomial-chaos expansion have very small magnitude. This implies that $\ell_1$-norm of all coefficients
\begin{equation}
\sum\limits_{|\basisInd|=0}^p{| c_{\basisInd} |} \approx \sum\limits_{|\basisInd|=0}^p{| \langle \ten{X}, \ten{W}_{\basisInd}\rangle |}
\end{equation}
should be very small.
\end{itemize}
  
  Combining the low-rank and sparse constraints together, we suggest the following optimization problem to compute $\ten{X}$ as an estimation (or approximation) of $\ten{Y}$:
 \begin{align}
\label{tensor_lrsp}
 \min_{\{\mathbf{u}_k^1, \cdots, \mathbf{u}_k^r \}_{k=1}^d} &\; \frac{1}{2} \| \mathbb{P}_{\Omega}\left(\sum\limits_{j=1}^r {\uvec_1^j \circ \cdots \circ \uvec_{\parNum}^j}  -{\ten{Y}}\right)\|_F^2 \nonumber \\
& +\lambda \sum\limits_{|\basisInd|=0}^p |\left\langle \sum\limits_{j=1}^r {\uvec_1^j \circ \cdots \circ \uvec_{\parNum}^j}, \ten{W}_{\basisInd}\right\rangle |.
\end{align}
In this formulation, we compute the vectors that describe the low-rank decomposition of $\ten{X}$. This treatment has a significant advantage: the number of unknown variables is $dqr$, which is only a linear function of parameter dimensionality $d$. We pick $\lambda$ and the size of $\Omega$ by empirical cross validations.

\subsubsection{Solve Problem (\ref{tensor_lrsp})} The optimization problem (\ref{tensor_lrsp}) is solved iteratively in our implementation. Specifically, starting from a provided initial guess of the low-rank factors $\{\mathbf{u}_k^1, \cdots, \mathbf{u}_k^r \}_{k=1}^d$, alternating minimization is performed recursively using the result of previous iteration as a new initial guess. Each iteration of alternating minimization consists of $d$ steps. At the $k$-th step, the $r$ vectors $\{\mathbf{u}_k^1, \cdots, \mathbf{u}_k^r \}$ corresponding to parameter $\xi_k$ are updated by keeping all other factors fixed and by solving (\ref{tensor_lrsp}) as a convex optimization problem.

\section{Numerical results}
In order to verify our tensor-recovery uncertainty quantification algorithm, we show the simulation results of two high-dimensional IC and MEMS examples. All codes are implemented in MATLAB and run on a Macbook with 2.5-GHz CPU and 16-G memory.
\begin{figure}[t]
	\centering
		\includegraphics[width=2.6in]{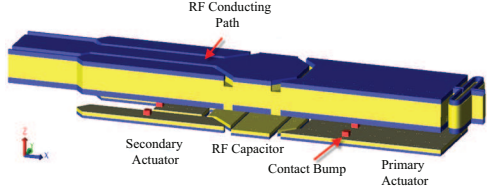} 
\caption{Schematic of the RF MEMS capacitor~\cite{zzhang:JMEMS2014}.}
	\label{fig:memsCap}
\end{figure}


\begin{table} [t]
\caption{Comparison of simulation cost for the MEMS capacitor.} 
\centering 
\begin{tabular}{c c c c} 
\hline
method & tensor product & sparse grid & proposed \\  \thickhline 

simulation samples & $ 8.9\times 10^{21}$ & $4512$ & $300$ \\
         
\hline 
\end{tabular} 
\label{table:mems_cost}
\end{table}

\subsection{MEMS Example (with 46 Random Parameters)}
We consider the MEMS device in Fig.~\ref{fig:memsCap}, and we intend to approximate its capacitance a $2$nd-order generalized polynomial-chaos expansion of $46$ process variations. As shown in Table~\ref{table:mems_cost}, using $3$ Gauss-quadrature points for each parameter, a tensor-product integration requires $3^{46}\approx 8.9\times 10^{21}$ simulation samples, and the Smolyak sparse-grid technique requires $4512$ simulation samples. 

\begin{figure}[t]
	\centering
		\includegraphics[width=3.7in]{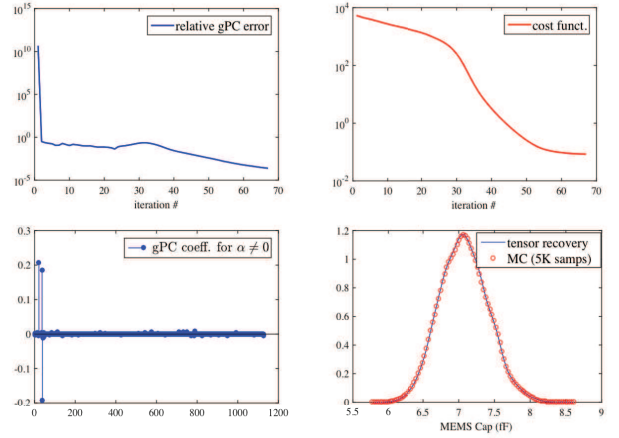} 
\caption{Numerical results of the MEMS capacitor, with $\lambda=0.01$. Top left: relative error of the generalized polynomial-chaos coefficients in iterations; top right: decrease of the cost function in (\ref{tensor_lrsp}); bottom left: sparsity of the obtained generalized polynomial-chaos expansion; bottom right: obtained probability density function compared with that from Monte Carlo. }
	\label{fig:mems_results}
\end{figure}

We simulate this device using only $300$ quadrature samples randomly selected from the tensor-product integration rules, then our tensor recovery method estimates the whole tensor $\ten{Y}$ [which contains all $3^{46}$ samples for the output $\out (\vecpar)$]. The relative approximation error for the whole tensor is about $0.1\%$ (measured by cross validation).  As shown in Fig.~\ref{fig:mems_results}, our optimization algorithm converges with less than $70$ iterations, and the generalized polynomial-chaos coefficients are obtained with a small relative error (below $10^{-4}$); the obtained model is very sparse, and the obtained density function of the MEMS capacitor is almost identical with that from Monte Carlo. Note that the number of repeated simulations in our algorithm is only about $1/4$ of the total number of basis functions. 

\subsection{7-Stage CMOS Ring Oscillator (with 57 Parameters)}
\begin{figure}[t]
	\centering
		\includegraphics[width=2.0in]{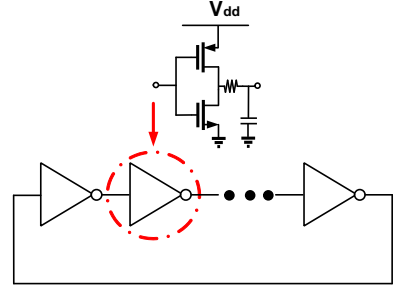} 
\caption{Schematic of the CMOS ring oscillator. }
	\label{fig:ring}
\end{figure}
We further consider the CMOS ring oscillator in Fig.~\ref{fig:ring}. This circuit has $57$ random parameters describing the variations of threshold voltages, gate-oxide thickness, and effective gate length/width. We intend to obtain a $2$nd-order polynomial-chaos expansion for its frequency by calling a periodic steady-state simulator repeatedly. The required number of simulations for different algorithms are listed in Table~\ref{table:ring_cost}, which clearly shows the superior efficiency of our approach for this example.
\begin{table} [t]
\caption{Comparison of simulation cost for the ring oscillator.} 
\centering 
\begin{tabular}{c c c c} 
\hline
method & tensor product & sparse grid & proposed \\  \thickhline 
simulation samples & $ 1.6\times 10^{27}$ & $6844$ & $500$ \\ \hline 
\end{tabular} 
\label{table:ring_cost}
\end{table} 

\begin{figure}[t]
	\centering
		\includegraphics[width=3.3in]{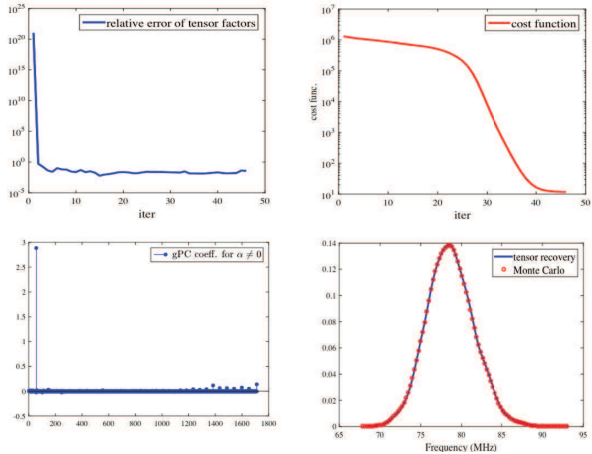} 
\caption{Numerical results of the ring oscillator, with $\lambda=0.1$. Top left: relative error of the tensor factors for each iteration; top right: decrease of the cost function in (\ref{tensor_lrsp}); bottom left: sparsity of the obtained generalized polynomial-chaos expansion; bottom right: obtained density function compared with that from Monte Carlo using $5000$ samples. }
	\label{fig:ring_results}
\end{figure}
We simulate this circuit using only $500$ samples randomly selected from the $3^{57} \approx 1.6\times 10^{27}$ tensor-product integration samples, then our algorithm estimates the whole tensor $\ten{Y}$ with a $1\%$ relative error.  As shown in Fig.~\ref{fig:ring_results}, our optimization algorithm converges after $46$ iterations, and the tensor factors are obtained with less than $1\%$ relative errors; the obtained model is very sparse, and the obtained density function of the oscillator frequency is almost identical with that from Monte Carlo. Note that the number of our simulations (i.e., $500$) is much smaller than the total number of basis functions (i.e., $1711$) in the generalized polynomial-chaos expansion.

\section{Conclusion}
This paper has presented a big-data approach for solving the challenging high-dimensional uncertainty quantification problem. Our key idea is to estimate the high-dimensional simulation data array from an extremely small subset of its samples. This idea has been described as a tensor-recovery model with low-rank and sparse constraints. Simulation results on integrated circuits and MEMS show that our algorithm can be easily applied to problems with over $50$ random parameters. Instead of using a huge number of (e.g., about $10^{27}$) quadrature samples, our algorithm requires only several hundreds which is even much smaller than the number of basis functions.


\section*{Acknowledgment}
This work was supported by the NSF NEEDS program and by AIM Photonics under Project MCE$\_$EPDA004. 

\bibliographystyle{IEEEtran}
\bibliography{SPI}

\end{document}